\documentstyle[11pt,aaspp4]{article}   % Nice, compact format w/out space jumps
\lefthead{Fabbiano \& Shapley }
\righthead{Multiwavelength emission properties of spiral galaxies}

\def\ergcm2s{erg cm$^{-2}$ s$^{-1}$} % ergs per cm**2 second
\def\etal{et al.}		% Present ApJ style, override older definitions
\def\n4038{NGC 4038/39}		% Use in text.  For title, use NGC 4038/4039
\def\rosat{{\it ROSAT}}		% ApJ uses Italics for 'ROSAT'
\begin{document}

\title{A Multivariate Statistical Analysis of Spiral Galaxy Luminosities.\\
II. Morphology-Dependent Multiwavelength Emission Properties }

\author{G. Fabbiano }
\affil{Harvard-Smithsonian Center for Astrophysics,\\
60 Garden Street, Cambridge, MA 02138}

\author{Alice Shapley}
\affil{California Institute of Technology,\\
Pasadena CA, 91125, USA}

% Notice that each of these authors has alternate affiliations, which
% are identified by the \altaffilmark after each name.  The actual alternate
% affiliation information is typeset in footnotes at the bottom of the
% first page, and the text itself is specified in \altaffiltext commands.
% There is a separate \altaffiltext for each alternate affiliation
% indicated above.
%
%\altaffiltext{1}{Visiting Astronomer, Cerro Tololo Inter-American Observatory. 
%CTIO is operated by AURA, Inc.\ under contract to the National Science
%Foundation.} 
%\altaffiltext{2}{Society of Fellows, Harvard University.} 

% The abstract environment prints out the receipt and acceptance dates
% if they are relevant for the journal style.  For the aasms style, they
% will print out as horizontal rules for the editorial staff to type
% on, so long as the author does not include \received and \accepted
% commands.  This should not be done, since \received and \accepted dates
% are not known to the author.

\begin{abstract}
This is the second of two papers based on a systematic multi-wavelength (X, B, H, 
12$\mu$m, FIR, 6cm) statistical analysis of the {\it Einstein Observatory Galaxy 
Catalog} sample 
of 234 `normal' S0/a-Irr galaxies. This sample is representative of 
spiral galaxies (Paper I), and its wide wavelength coverage provides a unique opportunity
for a systematic exploration of the relations among different emission bands, that
can give novel insight on the different emission processes and their relation to galaxian 
components, as a function of galaxy morphology.

We find clear morphological differences in the overall relationships among different
emission bands: while all wave-bands tend to be well correlated in late-type Sc-Irr galaxies,
in agreement with a general connection of all emission bands with the stellar component,
in bulge-dominant early-type S0/a-Sab galaxies, instead, our results point to
unrelated emission mechanisms
for the two groups of wave-bands  (12$\mu$m, FIR, 6cm) and (X, B, H).

A dependence on galaxy morphology is also found in the 12$\mu$m -- FIR link,
which was shown by the multivariate analysis of 
Paper I to be one the two strongest fundamental correlations in spiral galaxies, 
together with the well-known  B -- H correlation. 
We find that blue late-type dwarf galaxies appear to
lack 12$\mu$m emission, suggesting a lack of small-size dust grains in their ISM,
perhaps resulting from a particularly intense UV photon field. Also,
the IR -- B color-color plot of S0/a-Sab
galaxies presents more scatter than those of later-type galaxies, suggesting
both significant red giant circumstellar dust emission and nuclear emission.

Examining the well-known IR -- radio continuum link of spiral galaxies, 
we find surprising morphological differences: the `fundamental' radio-IR 
correlation flips from the mid-IR (12$\mu$m) emission in S0/a-Sab galaxies, to the FIR in 
Sc-Irr. These correlations also have different slopes:
the 6cm -- 12$\mu$m  correlation of S0/a-Sab galaxies
is steep and follows the same trend as that of the bright AGNs in the sample,
suggesting that low-luminosity AGN embedded in these `normal' 
galaxies may be responsible for the bulk of their mid-IR and radio emission;
the 6cm -- FIR correlation of Sc-Irr galaxies is linear, consistent 
with the well known star-formation link.

The X-ray emission is not strongly correlated with any of the other variables,
with the exception of the FIR in Sc-Irr, consistent with a general link 
with the young stellar population in these systems.
Contrary to preliminary reports, we find that
the X-ray -- B correlation is similarly non-linear (slope $\sim$1.5) in early-
and late-type spirals. We suggest that this 
apparent similarity stems from different processes in the two samples.
In S0/a-Sab's the steep slope may be due to the presence of some hot ISM in the more 
X-ray luminous systems, as observed in E and S0 galaxies. In Sc-Irr galaxies instead it
may be related to a luminosity-dependence of intrinsically absorbed X-ray
emission regions, connected with star formation activity. 
Obscured star-forming regions in higher  
luminosity galaxies can also explain other functional relations
of the correlations found in the Sc-Irr sample.

\end{abstract}

% The different journals have different requirements for keywords.  The
% keywords.apj file, found on aas.org in the pubs/aastex-misc directory, 
% contains a list of keywords used with the ApJ and Letters.  These are 
% usually assigned by the editor, but authors may include them in their 
% manuscripts if they wish. 


% That's it for the front matter.  On to the main body of the paper.
% We'll only put in tutorial remarks at the beginning of each section
% so you can see entire sections together.

% In the first two sections, you should notice the use of the LaTeX \cite
% command to identify citations.  The citations are tied to the
% reference list via symbolic KEYs.  We have chosen the first three
% characters of the first author's name plus the last two numeral of the
% year of publication.  The corresponding reference has a \bibitem
% command in the reference list below.
%
% Please see the AASTeX manual for a more complete discussion on how to make
% \cite-\bibitem work for you.   

\section{Introduction}

Understanding the structure, formation and evolution of galaxies is one
of the main themes of present-day astrophysics. This quest is made
difficult by the complexity of galaxies combined with our limited knowledge 
of their observational characteristics.
Different galaxian components [old, new, evolved stars; active nuclei (AGN);
the interstellar medium (ISM) -- gaseous; dusty; hot; cold] contribute in 
different amounts to the observed emission at different wavelengths, from
the radio to the X-rays 
(see \cite{gf90}). Therefore the comparison of global emission properties 
in a wide range of wavelengths can give us precious insight on the relative
importance of these components, as well as on the origin of some parts of the
emission spectrum. Since different emission bands have different sensitivities
to absorption, their comparison may also give us insight on the
dust content of the emitting regions (e.g. \cite{pal85}; \cite{ft87}).
Moreover, comparison of global multiwavelength emission properties of galaxies
of different morphology can give us insight on the relative presence of different
galaxian components throughout the Hubble sequence.

While most of the studies of galaxies
make use of individual energy bands, chiefly the optical, but also the radio,
and more recently the X-ray and infrared (IR), it is rarer to find work
comparing data from two or more emission windows. Yet, when this is done
interesting insights may follow. For example, the comparison of H-band and B-band 
photometry led to the discovery of the well known color-magnitude relation
for spiral galaxies (\cite{ahm79}; \cite{tma82}), a non-linear correlation between $L_B$
and $L_H$. The comparison of 
IRAS far-IR and radio continuum data led to the discovery of the well-known
strong correlation and to the convincing association of the radio continuum 
emission with the star-forming stellar population (\cite{dic84}; \cite{hel85};
\cite{dej85}); comparison of
CO, H$\alpha$ and IR data led to constraints on star formation efficiencies 
in spirals (e. g. \cite{y90}); comparison of multiwavelength data, including
X-rays, in late-type spirals suggested the prevalence of intrinsically obscured 
compact star-forming regions in higher luminosity galaxies (\cite{fgt88};
\cite{tfb89}).

This paper is a panchromatic study of a sample of 234 normal galaxies extracted
from the {\it Einstein Observatory Atlas and Catalog of Galaxies} (\cite{fkt92}, 
hereafter FKT). The sample is large enough to divide into morphological sub-samples.
Details on the sample selection and properties, and on the 
compilation of the multi-frequency data are given in Paper I (\cite{sfe98}), 
together with a description of the statistical analysis of these data.  Here we 
examine the astrophysical import of the results of Paper I, and
we support these results with further analysis. 
We take the panchromatic approach by comparing the global
emission properties of spiral and irregular galaxies in six emission
windows: X-rays (0.2-4~keV), optical (B), near- (H), mid- (12$\mu$m) and 
far-infrared (60-100$\mu$m, FIR), and radio continuum (6cm).
We also take into account explicitly the
morphological type of the galaxies, by dividing the sample in early
(S0/a-Sab), intermediate (Sb-Sbc) and late (Sc-Irr) subsamples, and intercomparing
the luminosity correlations in the three subsamples.

Previous panchromatic explorative work including the X-ray band was based on a 
much smaller sample of 51 spirals observed in X-rays with the {\it Einstein 
Observatory} (\cite{ft85}, hereafter FT; \cite{fgt88}
hereafter FGT; see \cite{f90}). This earlier work suggested interesting results
and motivated the present effort. 
Differences were reported in both scatter and power-law `slopes' of the
correlations, when comparing bulge-dominated and disk/arm-dominated
spirals, pointing to different contributions of bulge and disk to the various emission bands. 
In late-type spirals, all the correlations appeared to be very tight, but with 
different power-law slopes. Departures from linearity suggested the
prevalence of intrinsically obscured compact star-forming regions in higher 
luminosity galaxies. 
Based on the analysis of a yet smaller sample of 17 late-type spirals, 
the puzzling possibility was advanced of an intrinsic 
link between X-ray sources and cosmic ray sources responsible for the radio 
non-thermal continuum (FT, FGT). This radio -- X-ray link seems at odds with 
the prevalent association of radio continuum and the FIR-emitting star-forming 
component (e.g. \cite{hel85}). 

Some of the above reports may stem from limited 
statistics. Others may point to intrinsically complex astrophysical situations.
Our present sample, being nearly five times larger than the one used in previous
work, gives us an opportunity to look at some of these issues afresh with the
benefit of a much stronger statistical foundation.
Moreover, 
this sample comprises 58 S0/a-Sab galaxies, versus the 16 that
were included in the FT and FTG works. 
Furthermore, 
the inclusion of the 12$\mu$m band in the present analysis allows
for a more complete wavelength coverage than in previous work.
It therefore allows us for the first
time to study the average emission properties of early-type spirals with a
reasonable degree of confidence.

This paper is organized as follows: after the Introduction (\S 1.), \S 2. summarizes
the characteristics of the sample and the analysis methods used in Paper I;
\S 3.  discusses the evidence linking some emission bands prevalently to the disk/arm 
component and others to the bulge component as well; \S 4. and \S 5. discuss 
morphology-related
differences in the mid-, far-infrared, and radio continuum emission, and their 
implication for our understanding of the radio continuum and mid-IR emission
of S0/a-Sab galaxies; 
\S 6. discusses the X-ray -- B correlation and the implications of our results 
for the origin of the X-ray emission of bulge-dominant S0/a-Sab galaxies; 
\S 7. discusses the link of the X-ray emission of late-type spirals with the
star-forming stellar population, and \S 8. addresses the non-linear power-law
dependencies in disk/arm dominated galaxies; \S 9. summarizes the results.

\section{The Sample and Data Analysis }

The sample and data analysis procedures are fully described in Paper I.
The sample (extracted from FKT) consists of 234 `normal' spiral 
and irregular galaxies (from S0/a to Irr)
observed in (0.2-4)keV X-rays with the {\it Einstein Observatory}.
The FKT sample included also 21 bright Active Galactic Nuclei
(AGN), where the X-ray emission is clearly dominated by the nuclear source.
We do not use these galaxies in our analysis unless explicitly stated. 
As discussed in Paper I, the FKT sample selection is optical and
reflects the basic incompleteness
of the RC2. Since the sample is not statistically complete, nor X-ray selected,
it cannot be used to infer directly the X-ray luminosity function of spiral galaxies. 
However, the FKT sample offers a fair representation of spiral galaxies, with good coverage
in both morphological types and luminosity range, and it does not suffer from peculiar
distance biases (Paper I). It is therefore a good representative sample for 
studying average emission properties. This sample 
represents nearly a five-fold increase by comparison with previous 
statistical analyses of late-type galaxies observed in X-rays (FT; FGT).

As discussed in Paper I, multiwavelength fluxes were obtained through 
literature searches, and converted to luminosities for our analysis,
using the distances listed and discussed in Paper I.
The bands covered include, besides the X-rays, the optical B, near-IR
1.6$\mu$m H, IRAS 12$\mu$m and FIR (plus individual use in some cases
of the 60$\mu$m and 100$\mu$m data), and 6cm radio continuum.
The inclusion of 6cm, instead of 20cm, as an indicator of radio
continuum allows us to compare directly the results of the spiral sample 
with those of E and S0s (\cite{efk95}, hereafter EFK). 
In all cases, both detections and upper limits were included, to avoid 
obvious flux limit biases in the non-optical bands.

The data were analysed using the censored statistical analysis package
ASURV (e.g., \cite{iso86}; \cite{lav92}), and custom multivariate 
statistical  software developed by our group (Paper I). 
We tested all possible pairs of luminosity variables 
for statistically significant correlations, and also
compared several illustrative sets of flux ratios.  We then performed 
regression analysis on the luminosity correlations and derived regression 
`bisectors' representing the functional relations between any given pair of variables.
To discriminate between intrinsically strong correlations, and correlations
that may be the by-product of other stronger links, we performed a multivariate
conditional probability analysis using the Partial Spearman Rank method (\cite{ken76};
see FGT and EFK for previous applications).
The use of both detections and limits in the analysis, 
and the sample selection criteria
should protect us from distance biases. However,
we also included explicitly the distance in our multivariate analysis,
thus confirming that the correlations between luminosities are not
a spurious distance effect (see Paper I).

We analysed both the sample as a whole and three subsamples including galaxies
of different morphological types: the \underline{early sample} (58 
T=0-2 S0/a-Sab galaxies; we excluded from this subsample 7 Amorphous galaxies 
with T=0, see Paper I)
including galaxies with prominent bulges; the \underline{intermediate
sample} (61 galaxies; T=3-4; Sb-Sbc), typically consisting of luminous bulge/disk 
spirals; and the \underline{late sample} (108 galaxies; T=5-10; Sc-Irr), which 
includes galaxies dominated by the disk/arm star-forming component. 

\section{Morphological Effects in the Global (X-ray to radio)
Emission Properties of Spiral Galaxies}

It is not surprising that galaxies of different morphological types
may emit differently at different wavelengths, either because of their
stellar population mix, dust content, active star formation, properties of
the ISM, or size and accretion efficiency 
of the nuclear black hole. What we attempt in this 
work is a first systematic look at the multi-wavelength global emission properties of
spirals.
While our original motivation was to see how the X-ray emission fits in
the whole, and how it relates to other galaxian components and properties,
this work is not biased towards the X-ray properties, because of the 
properties of our sample (drawn from the RC2, and with comparable biases; Paper I), 
and our results ought to be valid for the overall panchromatic spectrum of spirals.
Comparison with established results will help establish how representative
our sample is. Our panchromatic approach may uncover new relations that may
shed light on some aspects of the global emission of spiral galaxies.

\subsection{Correlation Strengths}

In Paper I, we have found that both the significance of 
correlations between different emission bands and their functional
relationships (correlation `slope') vary with galaxy morphology.
Here, we will examine first the morphological dependence of the correlation `strengths',
we will discuss the slopes as well in the following sections of this paper.

Summarizing the results of Paper I:

\smallskip
\noindent
(1) {\bf All} correlations are very tight
(probability of chance correlation P$\leq 10^{-6}$) in the late (Sc-Irr) sample.

\smallskip
\noindent
(2) 
We notice two separate groups of inter-correlated
variables in the early (S0/a-Sab) sample: $L_{12\mu m}$, $L_{FIR}$, $L_{6cm}$ (all 
with P$\leq 10^{-6}$); and $L_X$, $L_B$, $L_H$. In the latter group $L_B$ is
strongly correlated with both $L_H$ and $L_X$ (P$\leq 10^{-6}$), while no 
significant correlation is found between $L_X$ and $L_H$.
Correlations between pairs of variables belonging to different groups are 
weaker or absent.

\smallskip
\noindent
(3)
The results for the intermediate sample (Sb-Sbc) are somewhat intermediate:
strong correlations persist among $L_{12\mu m}$, $L_{FIR}$, $L_{6cm}$ and
between $L_B$ and $L_H$, but significant links of $L_X$ with 
$L_{12\mu m}$, $L_{FIR}$, $L_{6cm}$ now appear.  

These results demonstrate the existence of  morphological dependencies in the
global emission properties of spiral galaxies,
and confirm the earlier reports of such effects (FT, FGT).
In late-type spirals and irregulars, the generally good correlations
point to a general connection of all emission bands with the stellar
component. In early-type bulge-dominant galaxies instead, the different
correlations strengths observed between the two groups of three variables
(X-ray, B, H) and (12$\mu$m, FIR, 6cm) suggest possibly unrelated emission mechanisms.

\subsection{Color Effects}

A closer look at the luminosity scatter diagrams (see Paper I for 
a full compilation) reveals that also the relative amount of emission (i.e., color)
in the two bands under examination may be morphology dependent. 
This effect is evidenced by the comparison of best-fit bisectors for
early and late samples, which are shown in two representative instances in
fig.~1. 

To quantify this effect for any given pair of variables,  we have used the
following approach: 1) Using the best-fit regression bisector
of the late-type (T=5-10) sample as a reference, we have built the
distributions of vertical residuals about this line,
for both the T=0-2 and the T=5-10 sample.
2) We have then compared these distributions, to see if we would detect
a systematic difference in their means, which would be symptomatic of a
morphological dependence of the flux ratios.
3) In order to gauge the significance of these differences, we have performed
two-sample tests on the two samples of residuals to derive the probability
that the two distributions are both drawn from the same parent population.
ASURV provides four different two-sample tests to use for censored data, which
differ in the way the censored data are weighted in calculating the test 
statistic. These methods are Gehan's Generalized Wilcoxon Test, the Logrank 
test, Peto and Peto's Generalized Wilcoxon, and Peto and Prentice's Generalized
Wilcoxon test. We have applied all these tests in each case. 
Fig.~2 shows representative histograms of residuals and Table~1
tabulates the results for all the distributions of residuals which could be tested, i. e.  
excluding correlations with double censoring.
A positive mean residual for the T=0-2 sample indicates that there is 
a relative excess in the Y-axis/X-axis ratio for this sample, when compared with the
T=5-10 sample. 

We see the well-known effect that early-type spirals tend to have redder colors
(larger H/B flux ratios) than later-type spirals, reflecting their older stellar
population (e.g., \cite{ahm79}). 
We find a similar H/X effect, while the ratio of X-ray and B fluxes
is not morphology dependent. This effect may suggest a closer link of the X-ray with the 
B-band emission (rather than with the H-band emission), in agreement with previous reports
(FT, FTG; see Paper I).
The morphological dependencies of the FIR, 6cm, and 12$\mu$m scatter diagrams 
cannot be tested formally, because of double censoring. However,
an inspection of the scatter diagrams in Paper I does not reveal any obvious
offsets in the early-type versus late-type correlations, although in some cases
the correlations follow different power-law relations. 
Relative to H, we find that early-type (bulge) galaxies tend to be underluminous
in FIR, 12$\mu$m and 6cm. A similar effect is seen when we compare FIR and 6cm
with B, while the B/12$\mu$m ratio does not appear to be morphology dependent.
Comparisons with the X-rays cannot be formally tested because of double censoring.
However visual inspection of the data  suggests a similiar behavior
as that observed with B.

The relative lack of radio and far-IR emission relative
to stellar and X-ray emission we observe in early-type spirals (S0/a-Sab), 
strengthens the possibility of unrelated emission mechanisms for these
two wavebands.

\subsection{Fundamental and Secondary Correlations}

Comparison between different correlations can be pushed further by using
conditional probability analysis (e.g. Spearman Partial Rank tests). These tests 
(see also Paper I) help in distinguishing between `fundamental' and
`secondary' correlations, by giving the probability of a correlation existing between 
two given variables, when other variables are held fixed. Fig.~3  shows 
results of these tests applied to all three morphological subsamples
and suggests some interesting trends. For these tests we have used all the
available points for each subset of three variables under consideration, 
thus maximizing the significance of the results for a given test. In Paper I instead
emphasis was given to the uniform analysis of the same sample observed in all 
six wavebands from radio to X-rays, resulting in smaller working samples. 

Summarizing the results displayed in fig.~3:

\noindent
(1) There are no significant morphological dependencies in the 
relations among X-ray, B, and H luminosities. The only fundamental link 
among these variables is the strong link between B and H (see Paper I),
which is present regardless of morphological type. 

\noindent
(2)
As reported in Paper I, we find a surprising morphological trend in the 
relations between 12$\mu$m, FIR and 6cm luminosities, that persists in 
a 3-way comparison, and is strenghtened by the present use of a larger
sample: (a)
While the FIR is always strongly correlated with the 12$\mu$m emission,
the connection between IR and radio continuum 
is morphology dependent. (b) There is a strong  fundamental link between 
12$\mu$m and 6cm emission in S0/a-Sab galaxies. (c) In Sc-Irr galaxies instead
the strong link of the 6cm emission is with the FIR; note that this 3-way
analysis results in a weaker and likely spurious anticorrelation between
6cm and 12$\mu$m. (d) In the intermediate
sample (Sb-Sbc), the result is transitional. The radio -- IR
link is weaker ($P \sim 1-2\%$) and it appears with both the 12$\mu$m
and the FIR. 

\noindent
(3)
The multivariate analysis confirms the morphology-dependent grouping of
the luminosity variables in two groups, disconnected in the S0/a-Sab sample 
but not in the late Sc-Irr sample (see \S3.1. above):
(X-ray, B, H) and (12$\mu$m, FIR, 6cm). All the many-variable Spearman Partial 
Rank tests tabulated in Paper I suggest no connections among these two
groups in early-type galaxies, 
while a significant link of the X-ray emission with the FIR emerges in most of the late-sample
tests. Three-variable tests including the FIR, X-ray  and 
either B or H show this effect clearly. The X-ray, H, FIR results are displayed
in fig.~3. 

In \S4, \S5 and \S6 we take a closer look  at these relations and discuss
further their astrophysical implications.

\section{The 12$\mu$m and FIR correlations}

The results of the Partial Rank analysis tabulated in Paper I show that the 
FIR -- 12$\mu$m correlation is one of the two strongest fundamental
correlations (the other
being H -- B), regardless of morphological type and of the number of 
variables considered in the analysis (see also fig.~3). We find that
the regression bisector slope of this correlation ($\alpha$, where
$L_{FIR} \propto L_{12\mu m}^{\alpha}$) flattens from early to 
late-type spirals (Paper I): 1.23$\pm$0.12 in S0/a-Sab, 
0.98$\pm$0.07 in Sb-Sbc (T=3-4), and 0.74$\pm$0.03 in Sc-Irr (T=5-10). 
The slopes of the early and intermediate samples are 
consistent with unity (a linear, everything increases at the same rate,
correlation) within the errors. However, the slope of  the late sample is 
significantly smaller than 1. Fig.~4 shows the scatter diagram for the 
late sample, with the regression bisector (solid line), and compares it
with the bisector for the early sample (dashed line).
The morphological dependency of the bisector
slopes may be at least in part a luminosity effect. In particular, the T=0-2 and T=3-4 
galaxies (which, within the errors, have consistent slopes) tend to be of high luminosity,
while T=5-10 galaxies include a significant number of low-luminosity systems. 
Omitting the T=5-10 galaxies with Log($L_{FIR}<42.5$),
the regression slope steepens becoming close to unity.
This result suggests that there may be intrinsic differences between
the production of FIR and 12$\mu$m in low- and high-luminosity late-type galaxies.

We have investigated further the 12$\mu$m -- FIR link by generating
12/100 vs. B/100 color-color plots for the three subsamples (fig.~5). In the T=3-4
sample, the distribution of points is essentially horizontal, consistent
with a constant 12/100 ratio. This is what one would expect if the 100$\mu$m and
12$\mu$m emission both arise from the ISM in spiral galaxies (e.g. \cite{hel91}),
with the 12$\mu$m emission resulting from the 
non-equilibrium heating of small size dust grains by the same UV photons 
that heat larger grains to give rise to the 100$\mu$m flux. As suggested by \cite{hel91}
the spread in the 12/100 ratios may result
from differences in the dust grain-size distribution of the ISM in different galaxies,
which may in part be linked to their star-formation activity. 
Smaller grains would be destroyed by the more intense UV photon field
of more active star-forming galaxies.
In agreement with a constant 12/100 color, the 12$\mu$m and FIR luminosities 
are linearly correlated in this sample. 
The analogous plot for T=5-10 galaxies 
shows a similar distribution except for a group of galaxies which appear to be 
both bluer (excess of B relative to 100$\mu$m) and brighter in the 100$\mu$m band 
relative to the 12$\mu$m than the bulk of the distribution
(numbered 1-7 in fig.~5). 
These objects belong to the group of low luminosity galaxies responsible for 
the flattening of the FIR -- 12$\mu$m correlation (fig.~4).
These blue dwarf galaxies may be undergoing active star formation, resulting
in a depletion of small size interstellar grains. Their blue colors suggests 
particularly intense UV emission.

The 12/100 vs. B/100 plot for T=0-2 (S0/a-Sab) galaxies differs from those of later
types in interesting ways. It presents larger scatter, with a 
number of points having larger 12/100 and/or B/100
ratios than in later-type galaxies. Some of these points are within the
locus of E and S0 galaxies (from EFK, solid polygon in fig.~5) which are
distributed mostly within a diagonal strip in this plane. This distribution suggests
that the 12$\mu$m emission in at least some S0/a-Sab galaxies is due
to photospheric and circumstellar emission of red giants [as in E and S0s,
(\cite{kna92}), where a strong correlation between 12$\mu$m and B emission
is observed (EFK)]. The points with Log (12/100)$<$0 to the left 
of the E and S0 swath could be explained with
extinction; however, their colors are also consistent with those of
later-type spirals, and may be indicative of interstellar emission.
There are also points in the 12/100 vs. B/100 plot of S0/a-Sab galaxies 
which lie in a region outside of the observed loci of either later
type spirals or E and S0s. Some of these galaxies are well known
Seyferts (which were excluded from our correlation analysis; they 
are identified in fig.~4 by circled symbols), 
suggesting that this region is occupied by luminous active nuclei.
There may be a cross-over of nuclear versus global galaxian emission
properties, depending on the relative luminosities of these components:
notice that galaxies with mini-AGNs (identified by enclosing squares,
from \cite{hfs97}) tend to be 
in the E and S0 locus, and follow the general distribution of T=3-4
galaxies as well. Note that the strong FIR-12$\mu$m correlations
of S0/a-Sab galaxies
would exist no matter what the energy source that gets re-radiated (stars or
AGN).

In summary, the results presented here show an overall connection 
between mid and far-IR emission in spiral galaxies, regardless of
morphological type, in agreement with earlier conclusions on the
interstellar origin for both IR emission bands (\cite{hel91}).
However, there are complications to this picture. Discrepancies
arise in bulge-dominated galaxies, where both red-giants and luminous AGN
emission may play a dominant role in the 12$\mu$m emission of some
galaxies; and in blue late-type dwarf galaxies, where star formation 
may be particularly active, resulting in the depletion of small size grains
and relative lack of 12$\mu$m emission. 

\section{The Infrared-6cm Connection: Star Formation and Nuclear Activity}

It is well known that radio continuum at 20cm and the  far-infrared are 
correlated. This correlation, discovered
soon after IRAS became operational (\cite{dic84}; \cite{hel85}; \cite{dej85}),
suggests a strong connection between non-thermal radio continuum emission in galaxies
and the star-forming stellar component. Its discovery resolved the debate on the
origin of the radio continuum emission of spirals (\cite{leq}; \cite{bf77};
\cite{h90}), by linking the production of
relativistic electrons to SNII in the young stellar population. 

As shown by our analysis (Paper I) the radio-continuum far-infrared correlation
is also highly significant at 6cm, both in early- and in late-type spirals.
However, when we explore the relative strengths of the correlations 
among 6cm, 12$\mu$m,  and FIR, we find intriguing significant morphological
differences (\S3, fig.~3) between early- and late-type spirals:
in S0/a-Sab  galaxies the 6cm radio continuum is directly
correlated with the 12$\mu$m rather than with the FIR; 
in Sc and later types (T=5-10) instead the 6cm radio continuum is   
directly correlated with the FIR emission.
These differences in the radio-continuum -- IR correlations suggest different
origins for the radio continuum emission of S0/a-Sab galaxies and later
type spirals.

Although the 6cm emission of late-type spirals differs from the non-thermal 20cm emission,
by including a fraction of thermal emission from HII regions
(\cite{ggk82}), the strong FIR -- 6cm link in these galaxies and the linearity
of this correlation ($L_{FIR} \propto L_{6}^{0.91 \pm 0.06}$),
are in agreement with the overall connection of radio emission and star formation,
derived from 20cm measurements.

The fundamental 6cm -- 12$\mu$m correlation of S0/a-Sab galaxies instead argues 
for a different origin of the 6cm radio continuum in these systems.
This  correlation appears steeper (slope = 
1.79$\pm$0.21) than the analogous one for the Sc-Irr sample, which
has a nearly linear slope of 0.88$\pm$0.07 (fig.~6), that agrees with the overall
star formation picture.
We suggest that nuclear activity may be responsible for the steep fundamental
correlation of S0/a-Sab galaxies.
As discussed in \S~4, the 12/100 - B/100 color-color plot of S0/a-Sab galaxies
suggests that nuclear activity may be important for their
12$\mu$m emission. Moreover, we find a remarkable similarity when we compare
the 6cm -- 12$\mu$m scatter plots of `normal' S0/a-Sab's and bright AGNs (fig~6a).
S0/a-Sab galaxies containing bright AGNs were not included in the regression
analysis (Paper I) but are plotted as circled points in the scatter diagram.
Their distribution is consistent with
that of the other points in the diagram, suggesting that the 12$\mu$m and 6cm
emission of all these galaxies may be connected with nuclear activity. 
If we calculate the regression bisector,
including the AGN, we find a very consistent result (slope $1.69 \pm 0.17$).
Note that a large number of the early sample galaxies host a low-luminosity 
AGN (see fig. 6a).

The connection of radio-continuum emission with the mid-IR, rather than with
the FIR is in itself suggestive of some kind of nuclear activity.
Seyfert galaxies and QSOs tend to have significant mid-IR emission (\cite{el94}).
In these bright AGNs the mid-IR emission may be due to thermal emission of
circumnuclear dust heated by the intense nuclear UV continuum.
Nonthermal radio and 10$\mu$m sources have also been found in the nuclei of spirals
galaxies, with remarkable spatial similarities (\cite{h89}).
In the latter type of nuclei shocks from supernovae
may collisionally heat small dust grains in the nuclear region
to $12 \mu$m temperatures, as well as accelerating electrons which
interact with the interstellar magnetic field to produce synchrotron radiation
(\cite{h89}).  Shocks would also be effective in enriching the ISM with small-size
grains, resulting in enhanced mid-IR emission. A similar picture may hold for
shocks connected with supersonic outflows from an active nucleus.

While the possibility of nuclear activity as an explanation of the radio --
mid-IR correlation in S0/a-Sab galaxies is attractive, the steep power-law
of this correlation cannot be easily explained in either of the above two
scenarios.  If the correlation were due to shocks resulting from phenomena
linked to star-formation, as in the \cite{h89} scenario,  we would
expect to see more of a proportionality between radio continuum and mid-IR
emission, as observed in late-type spirals. \cite{h89} report such
proportionality in point-to-point comparisons of their 10$\mu$m and radio
continuum maps. In the alternative AGN scenario, where re-radiation of
the UV continuum could be responsible for the mid-IR emission, the steep
6cm -- 12$\mu$m power law could be understood if there were a similar
steep dependency between the nuclear radio continuum and the higher
energy nuclear continuum. If we take the X-ray emission as an indicator
of the latter, and we assume that the UV continuum behaves accordingly,
we should then expect to find a similar steep radio -- X-ray correlation.
Except for a report by \cite{ung}, which however includes comparison of an
X-ray selected sample of active galaxies (the `Piccinotti' sample; \cite{pic}) 
with more powerful, optically/radio selected AGN, there is no evidence of 
such a trend. Within the 
Piccinotti sample, the radio -- X-ray luminosity correlation appears linear 
(\cite{ung}). The same is true for a correlation between radio core power 
and X-ray luminosity in powerful 3CR galaxies (\cite{f84}).

The above arguments suggest that if nuclear activity is responsible for the
6cm -- 12$\mu$m correlation of S0/a-Sab galaxies, these nuclei do not behave
either like those of more powerful AGN, or like nuclear regions where the
activity may be connected with particularly intense star formation (as in
\cite{h89}). However, a similarly steep correlation can be seen in relatively
radio-quiet E and S0 (EFK), where nuclear activity (either
non-thermal, e.g. \cite{fgt89}, EFK, \cite{ho98}, or possibly connected with nuclear
star formation, e.g. \cite{wro}) has also been raised as a possible explanation of
the radio continuum emission. If X-ray emitting hot halos are present (see \S~7)
in the more luminous/massive S0/a-Sab galaxies, these halos
may be responsible for diminishing their 
small-size dust grain population (e.g. see \cite{dwar}), resulting in a relatively 
less intense mid-IR emission at the higher luminosities.

We cannot exclude the possibility that S0/a galaxies alone may be responsible for the steep 
correlation because the slope flattens to $1.08 \pm 0.12$ if T=0 (S0/a) galaxies are excluded. 
However, this flattening may be due to the more restricted radio continuum luminosity
range covered by the T=1-2 galaxies in our sample (crosses in fig.~6a). In the region of
overlap, S0/a's cannot be distinguished from Sa-Sab's, suggesting no major
intrinsic difference. We do not find any substantial difference in the results
of the multivariate partial rank analysis if we exclude the S0/a galaxies. 
The 6cm-12$\mu$m
correlation is still the fundamental one. However, the sample size is in this case
reduced to only 14 galaxies. Further analysis of possible morphological differences
within early-type spirals will require assembling a substantially larger sample.

\section{The X-ray -- Star-Formation Link in Late-Type Spirals}

The connection between X-ray emission and star-formation activity was first suggested by
\cite{fabb82}, who reported higher X/B flux ratios in bluer
galaxies with disturbed morphologies, possibly resulting from a large number of
luminous young massive X-ray binaries, young SNR, and some hot ISM as well (e.g.
\cite{f96}). More recently, detailed multi-frequency high
resolution images are revealing one-to-one associations between bright X-ray emitting
regions and star-forming regions (e.g. in NGC4038/9, \cite{fsm97}, \cite{fab01}).
FGT (see also \cite{f90}) first discussed the X-ray -- FIR correlation in spiral 
galaxies and connected it with the star-forming galaxian component. Subsequently, 
\cite{djf} reported a connection between X-ray emission and FIR 
in the IRAS Bright Galaxy Sample.  This sample, however, includes early, intermediate, 
and late-type spirals, which, as we have now 
shown, differ in their integrated emission properties.

As discussed in Paper I, the results of the multivariate 
-- X-ray, B, H, 12$\mu$m, FIR, 6cm -- conditional probability analysis 
point to the  central role of star-formation in the global emission properties of
Sc-Irr galaxies. In these galaxies there are strong intrinsic correlations
of mid-IR and radio continuum with the FIR (Paper I). A possible, although weaker, link
of X-ray and FIR is also revealed in most of the tests in Paper I (see fig.~3).
Added support is provided by the existence of a significant
correlation of the $L_X/L_B$ luminosity ratio in late-type
spiral galaxies with the $L_{60}/L_{100}$ ratio (Paper I). The latter is a measure 
of the FIR color-temperature. Warm FIR color-temperatures
(higher $L_{60}/L_{100}$ ratios) have been  associated with star-formation
activity (e.g., \cite{hel91}). They have been found to occur {\it both} in high luminosity
spirals, suggesting the presence of hidden star forming regions in large disk galaxies, 
and in blue H$\alpha$-bright galaxies, where the association with star formation is 
more immediate (\cite{tfb89} (TFB) and refs. therein). Therefore,  the correlation of the $L_X/L_B$
excess with $L_{60}/L_{100}$ reinforces the link between $L_X/L_B$ and star-formation.

Table 2 summarizes the regression bisectors for the late sample (from Paper I)
and also gives bisectors obtained by omitting from the sample
the seven blue/FIR-bright dwarfs which appear a group apart in fig.~5.
The correlations between X-ray, FIR and 6cm luminosities all follow power-law slopes
near 1, consistent with a related origin through star formation, and with
the fact that these three emission bands are relatively little affected by 
extinction -- the X-rays are some, but much less than the optical band.
However, in the restricted higher luminosity sample
the X-ray - 6cm correlation departs significantly from linearity.
This flatter slope results from the omission of a single low-luminosity
detection from the sample  and as such the result calls for future
confirmation. If confirmed, it would be consistent with the X-ray -- radio-continuum 
slope found  by FGT. 

Based on a Partial Rank
analysis of a small sample of 17 T$ = 4 - 10$ galaxies, FGT suggested a
direct link between X-ray and radio continuum emission in late-type spirals.
Our present analysis does not confirm this suggestion.
If we analyse the correlations among X-ray, FIR and 6cm, with the 53 T$=5-10$ 
galaxies for which fluxes are available in all these three variables, we find 
that two correlations are significant, X -- FIR ($P = 0.007$) and FIR -- 6cm ($P < 0.005$),
the latter being by far the most significant. There is no direct statistical
link between X and 6cm.
Thus we conclude that the X-ray -- radio continuum correlation is likely
to arise from the connection of both emission bands with star-formation in
disk/arm dominated galaxies, rather to indicate a direct physical link
between X-ray and cosmic ray sources.

\section{The Steep X-ray -- B Correlation}

The $L_X - L_B$ correlation has been the subject of much attention in X-ray 
studies of normal galaxies of all morphological types (e.g. \cite{f89}, EFK), and it 
has been used to gain understanding on the nature and origin of the X-ray emission. 
However, our larger sample shows that this correlation is not a fundamental
one in spiral galaxies (Paper I). In early-type spirals there is
no strong fundamental link between X-rays and any of the other emission bands.
In late-type spirals there is a significant - but not overly strong - link 
only with the FIR (Paper I; \S~3; \S~6).
Perhaps, all the emphasis put into the study of this correlation is unwarranted.
However, even if this correlation is not the result of a direct causal link,
we can still learn by looking at it more closely.

We find (Paper I) that the $L_X - L_B$ relationship is significantly steeper than
linear in spirals, with slopes of 1.45$\pm$0.15 in S0/a-Sab galaxies, 1.53$\pm$0.17
in Sb-Sbc, and
1.48$\pm$0.13 in Sbc-Irr.  These results differ from previous reports of
a linear (power-law slope 1) relation between the X-ray and B-band
luminosities of spiral galaxies (FT; FGT), which 
suggested a link between the bulk of the X-ray emission 
and populations of evolved stellar sources (e.g. X-ray binaries, SNRs;
see Fabbiano 1989), belonging to the same stellar component responsible for the 
blue light. However, the earlier work  was based on a much smaller sample of 
galaxies than presently available (51 versus 234). 
While all these power-law slopes are statistically
consistent with each other, there are considerations that support an intrinsic
difference in the phenomena responsible for the bulk of the X-ray emission in early-
and late-type spirals. We have discussed in \S~6 the
linear X-ray -- FIR correlation of the late sample, which suggests
that the non-linear X-ray -- B correlation of late-type
spirals and irregulars may arise from extinction in star-forming complexes in
the most luminous galaxies. Extinction would affect the B band more than the X-rays,
causing a deviation from linearity (\S~8). This explanation does not
apply easily to early-type spirals, where dusty active star-forming regions
are rare.

The nature of the X-ray emission of early-type, prominent-bulge spirals has been
the subject of an on-going controversy, which has sought to establish if and how
much of this emission can be ascribed to the thermal emission of an optically thin hot 
gaseous medium, that may be used to measure the mass of the parent galaxy 
(\cite{fjt}, FT, \cite{tf85}, FGT, \cite{kft}, \cite{fj97},
\cite{ser}), in addition to the X-ray binary population that must be present (e.g. 
\cite{f89}) and it is now seen directly in Chandra images (e.g. \cite{ir00}). 
Detailed hydrodynamical models developed for S0 galaxies may apply, and suggest
that although large static halos are not expected, hot ISM in a complex flow state 
may be found (\cite{dec98}).
The steep X-ray -- B correlation may suggest the presence of gaseous halos
in X-ray luminous S0/a-Sab galaxies, by
analogy with the similarly steep correlation observed in E and S0 galaxies (\cite{tf85}, 
\cite{cft}, EFK, see also the ASCA results of \cite{mat97}).
Supporting evidence is provided  by X-ray spectral data.
The average {\it Einstein} X-ray colors of Sa galaxies are similar 
to the colors of S0 and E galaxies where hot halos are found, 
rather than to those of later types, which
are characterized by harder X-ray spectra, typical of X-ray binaries (\cite{kft}). 
ASCA higher resolution CCD spectra of NGC~4594 (\cite{ser}) and 
NGC~4736 (\cite{rob99}) also support the presence of a
thermal gaseous component in these galaxies, which is confirmed
in NGC~4736 by the high spatial/spectral resolution Chandra ACIS-S image
(\cite{pell01}; another recent example of a Chandra observation that shows both
a population of point-like sources and diffuse emission is the S0/a NGC~1291
\cite{ir00}).

`Hidden' AGNs may in principle add to the X-ray emission and cause a
steep X-ray -- correlation. However, we believe that this is unlikely.
Fig.~7 shows that luminous AGNs indeed have very significant X-ray excess luminosity, but 
their distribution is not consistent with the high luminosity extrapolation of 
the X-ray -- B correlation. A large number of galaxies in this sample host mini-AGNs
(\cite{hfs97}, see fig~7), but it is unlikely that the X-ray luminosity of these 
galaxies is dominated by this emission. This conclusion is supported by X-ray imaging for at
least two X-ray luminous galaxies:  NGC~4594 (\cite{fj97}, \cite{ho01};
a LINER galaxy with a compact nuclear radio
source, \cite{hec80}, \cite{con82}) and NGC~4736 (M94; \cite{cui}, 
\cite{rob99}, \cite{pell01}); in both galaxies 
the nucleus does not dominate the emission in the soft X-ray band. 
In NGC~4736, in particular, the Chandra image shows that the nucleus is not
even the most luminous source in the central region of this galaxy (\cite{pell01}).
Moreover, a recent high resolution Chandra survey of a 
number of mini-AGN galaxies shows in the majority of them nuclear sources
with luminosities well below the total galaxian emission (\cite{ho01}).

\section{Non-Linear Power-Law Dependencies in Disk/Arm-Dominated Spirals:
extinction and other effects}

We have mentioned earlier in this paper and shown in Paper I that in the 
late (T=5-10; Sc-Irr) sample all the correlations are highly significant. We have also
discussed how some of these correlations (e.g. X-ray -- FIR) are close to 
a power-law of slope 1.  Others (e.g. B -- H, X-ray -- B) are not (Table 2).

Extinction in dusty star-forming regions 
may be responsible for some of these departures from linearity.
Extinction would affect the
B emission more than longer wavelengths such as H, 12$\mu$m, FIR, 6cm, 
or higher energies such as the X-rays. 
The $L_B \propto L_{FIR}^{0.7}$ relation was first observed by FGT, and explained 
(see also TFB) with an enhanced dusty compact starburst component 
in higher-luminosity galaxies.  This scenario would also account for the
flat $L_B - L_{6cm}$ power-law: the longer wavelength
radio-emission from the star-burst component would not be obscured
by the dust in these star-forming regions, and so would scale linearly
with the FIR and according to the observed power-law with the optical.
The X-ray -- B relationship is also not linear,
and it could be consistent with the obscuration picture.
Assuming 2.5 magnitudes of visual extinction, which is the average value
required to explain the departure from linearity of the FIR -- B correlation
(FGT), we find that the corresponding extinction in the X-ray emission would only be
$A_X \leq 0.5$~mag, much less than the extinction in B. Although $A_V=2.5$~mag 
exceeds the average galaxian
extinction, a large amount of obscuration is present in localized star-forming
regions (e.g. \cite{errl}). A comparison
of X-ray and other waveband emissions in the Antennae galaxy (NGC~4038/9; \cite{fsm97})
supports the above picture. Intense X-ray, radio, and IR emission regions are 
co-spatial with the HII regions and nuclear region of this merger galaxy.
Considering that the H-band emission is significantly enhanced by red supergiants in 
dusty star-forming regions (e.g. \cite{errl}), the same star-formation/extinction 
scenario may naturally explain the flat power-law $L_B \propto L_H^{0.7}$, 
discovered by \cite{ahm79}.

The FIR -- 12$\mu$m departure from linearity has been discussed in \S 4 and 
ascribed at least in part to the effect of low-luminosity blue galaxies with large 
FIR/12$\mu$m ratios. However, Table 2 shows that after these galaxies are removed
the correlation is still flatter than linear.
This flat FIR-12$\mu$m correlation may be responsible
for the flatter relations of B (and X-rays) with 12$\mu$m than with FIR.
A possible explanation for this residual effect -- to be checked with future IR data --
is that our sample is still contaminated by some `blue dwarf' galaxies at the 
low-luminosity end.

\section{Summary and Conclusions}

We have performed (Paper I) a statistical analysis of 234 normal spiral galaxies 
from the sample of FKT, which were all
observed in X-rays with the {\it Einstein Observatory}.
This analysis explores relations among 6 emission bands, including
X-rays (0.2-4~keV), B, H, 12$\mu$m, FIR, and 6cm, in the entire sample of 234
galaxies and in subsamples comprising S0/a-Sab, Sb-Sbc,
and Sc-Irr galaxies respectively. In this paper
we report morphology-related effects in these results.
We also explore the implications of our results for the emission 
processes of early- (S0/a-Sab) and late-type (Sc-Irr) galaxies separately. 
The main conclusions of this paper are summarized below:

\noindent
(1) Bivariate and multivariate correlation analyses suggest clear
morphological dependencies of the global radio-IR-optical-X-ray
emission properties:

\noindent
-- in late-type galaxies (Sc-Irr), all the emission
properties are correlated suggesting a broad connection to the galaxian stellar population;

\noindent
-- in early-type spirals (S0/a-Sab) instead, 12$\mu$m, FIR, and 6cm emissions 
are not connected with either the stellar population responsible for the
global near-IR and optical light (H and B) or with the X-ray emission.

\noindent
-- the intermediate sample of Sb-Sbc (T=3-4) galaxies presents intermediate
properties, with some links appearing between X-rays and FIR.

\noindent
(2) The strong, always present, fundamental 12$\mu$m and FIR correlation (see
also Paper~I) agrees with the picture that
both emission bands are due to the same photon field interacting with 
different size grains (\cite{hel91}). However, we also find some effects that point
to morphology-related differences in the  IR emission processes, and possibly
to different (stellar and AGN) sources of radiation in different types of
galaxies:

\noindent
-- The 12/100 vs. B/100 color-color plot
of  S0/a-Sab galaxies presents more scatter than those of later 
morphological types, that could be explained with both significant
red giant circumstellar dust (as in ellipticals and S0s, \cite{kna92}), and nuclear
emission. 

\noindent
-- Blue late-type dwarf galaxies appear to be underluminous in
the mid-IR, suggesting that the composition of the ISM in these galaxies
lacks the small grains responsible for the mid-IR emission. This 
difference may be related to the more intense UV field, suggested by 
their blue colors.

\noindent
(3) Our results suggest a difference in the
production of radio continuum and IR emission in early- and late-type
spirals.

\noindent
--  In Sc-Irr galaxies we find a fundamental, strong, linear radio -- FIR correlation,
consistent with the previously reported connection of the radio continuum emission 
with the star-forming stellar population (e.g., \cite{hel85}). 

\noindent
--  In S0/a-Sab galaxies instead the fundamental radio -- infrared link is the
6cm -- 12$\mu$m correlation, and this correlation is steeper
than linear ($L_{6cm} \propto L_{12}^{1.79 \pm 0.21}$). We propose
nuclear activity as the source of
this correlation: the close link with the mid-IR argues in itself for this
possibility, since mid-IR emission is enhanced in AGN (\cite{el94}); 
moreover, the correlation is also consistent with the distribution of bright
AGNs in the same diagram.
The power-law slope of this correlation is however difficult to explain. Linear
correlations would be expected both in the picture of shock-related nuclear emission
(e.g., \cite{h89}), and in the more traditional AGN scenario, where the IR emission
may be due to re-radiation of the nuclear continuum. The observed non-linearity
may be related to the presence of hot gaseous halos in the more luminous galaxies,
which may result in depletion of small-size dust grains (and thus a decrease of the
12$\mu$ flux). A similar mechanism may operate in E and S0 galaxies, where
a steep radio -- IR correlation has also been reported (EFK).

\noindent
(4) In Sc-Irr galaxies the strongest link of the X-ray emission with a stellar
indicator is the linear X-ray -- FIR correlation.
Conditional probability analysis shows that this is
a far more significant link that the X-ray  -- B one, suggesting a 
connection of the X-ray emission with the star forming stellar component.
This conclusion is reinforced by the $L_X/L_B - L_{60}/L_{100}$ correlation
found in Paper I, which associates more intense X-ray emission with hotter
IR colors.
We find no evidence of a strong intrinsic link of X-ray and radio continuum
emission, contrary to the early suggestion of FGT.

\noindent
(5) Contrary to previous reports, based on much smaller samples (FT, FGT), the
X-ray -- B correlation is steeper than linear ($L_X \sim L_B^{1.5}$), regardless
of morphological type. 

\noindent
-- In the light of the stronger linear X-ray -- FIR correlation,
in Sc-Irr this slope may be explained with absorption of the 
blue light in dusty intense X-ray emitting star-forming regions in bright galaxies.

\noindent
-- In S0/a-Sab galaxies it suggests the presence of an additional emission component,
dominating the X-ray emission of the more luminous galaxies.
Comparisons with the analogous correlation observed in E and S0s,
and with the distribution of AGNs in the $L_X - L_B$ plot, suggest
that hot ISM in X-ray brighter galaxies (as in E and S0s), rather than
nuclear activity may be the more viable explanation.

\noindent
(6) Finally, examining the power-law slopes of the correlations in the late-type spirals 
(T=5-10), we find linear and non-linear correlations. These different slopes may be 
related to the prevalence of obscured starforming regions in higher luminosity
galaxies, as discussed in FGT.

Given the characteristics of our sample and the analysis techniques employed
we believe these results to be free from strong biases (Paper I), and therefore
to be representative of the characteristics of normal spiral galaxies.
This conclusion is reinforced by the general agreement with previous IR-based
work by other authors, using different representative samples of spiral 
galaxies. However, to confirm and expand these results, the assembly of
larger, well defined, possibly complete, samples of spiral galaxies with 
multi-wavelength coverage is needed.
Our results suggest many lines of future investigations:
a re-examination of the radio continuum -- IR correlations and their morphological
dependencies; high resolution 
observations of early-type spirals in the X-rays, mid-IR, and radio continuum, 
to isolate nuclear sources in these bands, and to quantify the amount of hot
ISM in the X-rays; comparison with optical emission line 
properties, to further explore the presence and properties
of low-luminosity active nuclei; multi-wavelength imaging of luminous late-type
galaxies to confirm the impact of dusty star-forming regions in their overall
emission.

% We show the use of several of the displayed math environments described
% in the User Guide, and you get a healthy dose of mathematical typesetting
% examples.  Also, observe the use of the LaTeX \label command after the
% \subsection to give a symbolic KEY to the subsection for cross-referencing
% in a \ref command.  LaTeX automatically numbers the sections, equations,
% tables, etc., as it goes, so in general you don't know what number something
% is going to have.  We'll refer to the "hairymath" section a little later.

\acknowledgments

We thank Paul Eskridge and Martin Elvis for comments and Steve Willner for
discussions on the near-IR emission.
This work was supported by NASA grants NAG~5--2281 and NAG~5--1937
(\rosat); NAGW--2681 (LTSA); and by NASA contract NAS~8--39073 (Chandra
X-ray Center).

\clearpage

% Camera-ready tables, produced with either the apjpt4 or aj_pt4 style files,
% can be referenced within a table environment using \dummytable.  This acts
% like a place holder and bumps the table counter.   For this particular
% manuscript, tab3 refers to the table in file samp2tbl.tex.

%\begin{table}
%\dummytable\label{tab3}
%\end{table}

% This is the last table for this paper (as well as the first), so we
% should follow it with a \clearpage.  In order to force all the floating
% tables out of their buffers and onto vertical page lists, we must use
% \clearpage rather than \newpage. 

\clearpage

\clearpage

% Following is an example:
% \figcaption[sgi9259.eps]{This is the first figure and it uses sgi9259.eps as
% its EPS figure file. \label{fig1}}

\figcaption{Scatter plots of $L_X - L_{FIR}$ and $L_H - L_{FIR}$ for the early (T=0-2, 
S0/a-Sab only) and late (T=5-10, Sc-Irr) samples. The solid lines represent the 
best-fit regression bisectors. The dashed lines on the early sample plots represent 
the corresponding
late samples regression bisectors. Filled squares represent galaxies detected in both
axes; circles represent upper limits in both axes; left-pointing triangles represent
X-axis upper limits; down-pointing triangles represent Y-axis upper limits;
squares surrounding other symbols are galaxies with mini-AGNs (\cite{hfs97})}

\figcaption{Histograms of residuals (in $L_X$) relative to the best-fit regression
bisector of the $L_X - L_{FIR}$ correlation of the late (T=5-10, Sc-Irr) sample.
Top: for the late sample. Bottom: for the early (T=0-2; S0/a-Sab) sample.}

\figcaption{ Summary of 3-variables Partial Rank results:
B, H, X-ray (top), FIR, 12$\mu$m, 6cm (middle), and X-ray, 
H, FIR) (bottom), for the three morphological subsamples. Test statistics and
probabilities (see Paper I) are given for each correlation. 
Two connecting lines represent very significant 
correlations ($P < 0.005$); one connecting line represents significant correlations
$P\leq 0.03$); and no line represents a lack of significant correlation.
The number of objects included in the test is given at the center of each diagram.
}

\figcaption{FIR -- 12$\mu$m scatter diagram for Sc-Irr galaxies (T=5-10).
Same symbols as in fig.~1, except for numbers
identifying the blue dwarfs of fig.~4. The solid line 
is the regression bisector for the entire late-sample fit. 
The dotted line is the bisector of the early-sample.}

\figcaption{12/100 vs. B/100 plot for the early (T=0-2), intermediate (T=3-4),
and late (T=5-10) samples. Bright AGN points are surrounded by circles; these were
not included in the correlation analysis. Triangles represent upper limits. 
Squares surrounding other symbols are galaxies with mini-AGNs (\cite{hfs97}).
The area outlined in the T=0-2 plot identifies the region occupied by S0 (from EFK).
Blue FIR-bright dwarves are labelled in the T=5-10 plot: 1=IC~1613; 2=NGC~4236;
3=NGC~247; 4=NGC~4244; 5=SMC; 6=IC~2574; 7=NGC~6822. }

\figcaption{ $L_{6cm} - L{12}$ correlation for S0/a-Sab (a), and 
Sc-Irr (b) galaxies.  Filled squares are detections in both 
variables, downward pointing triangles are upper limits in $L_{6cm}$, leftward 
pointing triangles are upper limits in $L_{12 }$, and
circles are limits in both variables.  Circled points in fig. (a) represent
known bright AGN, which were not included in the fit; crosses represent S0/a
galaxies.  Diagonal lines represent the best-fit regression bisectors.
Squares surrounding other symbols are galaxies with mini-AGNs (\cite{hfs97}).
}

\figcaption{ $L_X - L_B$ diagram for S0/a-Sab galaxies.
Filled squares are detections, triangles are upper limits
in $L_X$. The diagonal lines represent the best-fit regression bisectors.
The circled points at the top of the S0/a-Sab scatter diagram represent
known bright AGN, which were not included in the fit. 
Squares surrounding other symbols are galaxies with mini-AGNs (\cite{hfs97}).}


\begin{thebibliography}{}



\bibitem[Aaronson, Huchra \& Mould 1979]{ahm79}Aaronson M., Huchra, J., 
Mould, J. 1979, \apj, 229, 1. 

\bibitem[Biermann \& Friecke 1977]{bf77}Biermann, P. \& Friecke, K. 1977,
\aap, 54, 461.

\bibitem[Canizares, Fabbiano \& Trinchieri 1987]{cft}
Canizares, C. R., Fabbiano, G., Trinchieri, G. 1987, \apj, 312, 503

\bibitem[Cui \etal\ 1997]{cui} Cui, W., Feldkhun, D., Braun, R. 1997, \apj, 477, 693

%\bibitem[Charles \& Seward 1995]{cs95}Charles, P. A., Seward, F. D. 1995, `Exploring
%the X-ray Universe' (Cambridge: University Press)

\bibitem[Condon \etal\ 1982]{con82}Condon, J.J., Condon, M.A., Gisler, G., Purcell, J. J.
1982, \apj, 252, 102



\bibitem[de Jong \etal\ 1985]{dej85}de Jong, T., Klein, U., Wielebinski, R., Wunderlich, E.
1985, \aap, 147, L6

\bibitem[D'Ercole \& Ciotti 1998]{dec98}D'Ercole, A. \& Ciotti, L. 1998, \apj, 494, 535

\bibitem[de Vaucouleurs, de Vaucouleurs \& Corwin 1976]{RC2}de Vaucouleurs, G., 
de Vaucouleurs A., Corwin, H.G. 1976,
Second Reference Catalogue of Bright Galaxies (Austin:
University of Texas Press)

\bibitem[David \etal\ 1992]{djf}David, L. P., Jones, C., Forman, W. 1992, \apj, 388, 82

%\bibitem[Devereux \& Eales 1989]{dev89}Devereux, N., Eales, S. 1989, \apj, 340, 
%708

\bibitem[Dickey \& Salpeter 1984]{dic84}Dickey, J.M., Salpeter, E. E. 1984, \aj,
 284, 461


\bibitem[Dwek \& Arendt 1992]{dwar}Dwek, E., Arendt, R.G., 1992, Ann.Rev.A.Ap., 30, 11


\bibitem[Elvis \etal\ 1994]{el94}Elvis, M.,Wilkes, B. J., McDowell, J. C., 
Green, R. F., Bechtold, J., Willner, S. P., Oey, M. S., Polomski, E., 
Cutri, R. 1994, \apjs, 95, 1

\bibitem[Engelbracht \etal\ 1996]{errl}Engelbracht, C. W., Rieke, M.J., Rieke, G. H., Latter, W. B. 1996,
Ap. J., 467, 227.


\bibitem[Eskridge, Fabbiano \& Kim 1995]{efk95}Eskridge, P.B., Fabbiano, G., 
Kim, D.-W. 1995, \apjs, 97, 141 [EFK]

\bibitem[Fabbiano 1989]{f89}Fabbiano, G. 1989, Ann.Rev.A.Ap., 27, 87.

\bibitem[Fabbiano 1990]{f90}Fabbiano, G. 1990 in {\it Windows on Galaxies}, 
eds. G. Fabbiano,
J.S. Gallagher, A. Renzini, p. 231. Dordrecht: Kluwer

\bibitem[Fabbiano 1996]{f96} Fabbiano, G. 1996, in R\"ontgenstrahlung
from the Universe, ed.\ H. U. Zimmermann, J. E. Truemper \& H. Yorke,
MPE Report 263, p.\ 347

\bibitem[Fabbiano \etal\ 1982]{fabb82} Fabbiano, G., Feigelson, E., \&
Zamorani, G. 1982, \apj, 256, 397

\bibitem[Fabbiano \& Juda 1997]{fj97}Fabbiano, G., and Juda, J.Z. 1997, Ap. J., 476,
666


\bibitem[Fabbiano \& Trinchieri 1985]{ft85}Fabbiano, G., Trinchieri, G. 1985, \apj, 296, 430
(FT)

\bibitem[Fabbiano \& Trinchieri 1987]{ft87}Fabbiano, G., Trinchieri, G. 1987, \apj, 315, 46.

\bibitem[Fabbiano, Gioia \& Trinchieri 1988]{fgt88}Fabbiano, G., Gioia, 
I.M, Trinchieri, G. 1988, \apj, 324, 749 (FGT)

\bibitem[Fabbiano, Gioia \& Trinchieri 1989]{fgt89}Fabbiano, G., Gioia,
I.M, Trinchieri, G. 1989, \apj, 347, 127



\bibitem[Fabbiano, Kim \& Trinchieri 1992]{fkt92}
Fabbiano, G., Kim, D.-W., Trinchieri, G. 1992, \apjs, 80, 531.
(FKT)

\bibitem[Fabbiano \etal\  1984]{f84}Fabbiano, G., Miller, L., Trinchieri, G.,
Longair, M., Elvis, M. 1984, \apj, 277, 115.

%\bibitem[Fabbiano, Trinchieri \& Van Speybroeck 1987]{ftv}Fabbiano, G., Trinchieri, G.,
%Van Speybroeck, L.S. 1987, \apj, 316, 127



\bibitem[Fabbiano \etal\ 1997]{fsm97}Fabbiano, G., Schweizer, F., Mackie, G. 
1997, \apj, 478, 542

\bibitem[Fabbiano, Zezas \& Murray 2001]{fab01} Fabbiano, G., Zezas, A., \& 
Murray, S. S. 2001, \apj, 554, 1035

\bibitem[Forman, Jones \& Tucker 1985]{fjt}Forman, W., Jones, C., Tucker, W. 1985, \apj, 293, 102

\bibitem[Gallagher \& Fabbiano 1990]{gf90}Gallagher, J, Fabbiano, G. 1990, 
in {\it Windows on Galaxies}, eds. G. Fabbiano, J.S. Gallagher, A. Renzini, p.1.
Dordrecht: Kluwer

\bibitem[Gioia, Gregorini \& Klein 1982]{ggk82}Gioia, I.M., Gregorini, L., Klein, U.
1982, \aap, 116, 164

\bibitem[Gunn, Stryker \& Tinsley 1981]{gst81} Gunn, J.E., Stryker, L.L., Tinsley, 
B. M. 1981, \apj, 249, 48

\bibitem[Heckman 1980]{hec80} Heckman, T.M. 1980, \aap, 87, 152


%\bibitem[Helou \& Bicay 1993]{hb93}Helou, G., Bicay, M. D. 1993, \apj, 415, 93

\bibitem[Helou, Soifer \& Rowan-Robinson 1985]{hel85}
Helou, G., Soifer, B.T., Rowan-Robinson, M. 1985, \apj, 298, L7. 

\bibitem[Helou, Ryter \& Soifer 1991]{hel91}Helou, G., Ryter, C., Soifer, B.T. 1991,
\apj, 376, 505.

\bibitem[Ho, Filippenko \& Sargent 1997]{hfs97}
Ho, L. C., Filippenko, A. V., \& Sargent, W. L. W. 1997, \apjs, 112, 315


\bibitem[Ho 1999]{ho98} Ho, L. C. 1999, \apj, 510, 631

\bibitem[Ho \etal\ 2001]{ho01} Ho, L. C. et al 2001, \apj (letters), 549, L51

\bibitem[Ho \etal\ 1989]{h89} Ho, P. T., Turner, J. L., Fazio, G., Willner, S. P. 1989,
\apj, 344, 135


\bibitem[Hummel 1980]{hum}Hummel, E. 1980, Ph. D. Thesis, Univ. of Groningen, 
Nederlands


\bibitem[Hummel 1990]{h90}Hummel, E. 1990,
in {\it Windows on Galaxies}, eds. G. Fabbiano, J.S. Gallagher, A. Renzini, p. 141,
Dordrecht: Kluwer


\bibitem[Irwin, Bregman \& Sarazin 2000]{ir00}Irwin, J. A., Bregman, J. N., 
Sarazin, C. L. 2000, AAS, 197111.14

\bibitem[Isobe, Feigelson \& Nelson 1986]{iso86}
Isobe, T., Feigelson, E.D., Nelson, P.I. 1986, \apj, 306, 490.


\bibitem[Kendall \& Stuart 1976]{ken76}Kendall, M., Stuart, A. 1976, The Advanced Theory
of Statistics, Vol. 2 (New York: Macmillan)


\bibitem[Kim, Fabbiano \& Trinchieri 1992]{kft}
Kim, D.-W., Fabbiano, G., Trinchieri, G. 1992, \apj, 393, 134



\bibitem[Knapp \etal\ 1992]{kna92}Knapp, G.R., Gunn, J.E., Wynn-Williams, C.G. 1992, 
\apj, 399, 76




\bibitem[LaValley, Isobe \& Feigelson 1992]{lav92}
LaValley, M.P., Isobe, T., Feigelson, E.D. 1992, BAAS, 24, 839.


\bibitem[Lequeux 1971]{leq}Lequeux, J. 1971, \aap, 15, 42.

\bibitem[Matsumoto \etal\ 1997]{mat97}Matsumoto, H., Koyama, K., Awaki, H., Tsuru, T.,
Loewenstein, M., Matsushita, K. 1997, \apj, 482, 133

\bibitem[Palumbo \etal\ 1985]{pal85}Palumbo, G.G.C., Fabbiano, G., Fransson, C.,
Trinchieri, G. 1985, \apj, 298, 259.

\bibitem[Pellegrini \etal\ 2001]{pell01}Pellegrini, S. et al 2001, in preparation

%\bibitem[Persson \& Helou 1987]{ph87}Persson, C. J. and Helou, G. 1987, \apj, 314, 513.

\bibitem[Piccinotti \etal\ 1982]{pic} Piccinotti, G., Mushotzky, R.F., Boldt, E.A.,
Holt, S.S., Marshall, F.E., Serlemitsos, P.J., Shafer, R.A. 1982, \apj, 253, 485

\bibitem[Roberts \etal\ 1991]{retal}Roberts, M.S., Hogg, D.E., Bregman, J.N., 
Forman, W.R., Jones, C. 1991, \apjs, 75, 751

\bibitem[Roberts, Warwick \& Ohashi 1999]{rob99}Roberts, T. P., Warwick, R. S.
\& Ohashi, T. 1999, MNRAS, 304, 52

\bibitem[Shapley, Fabbiano \& Eskridge 2001]{sfe98}Shapley, A., Fabbiano, G.,
Eskridge, P. B. 2001, \apjs, in press (Paper I).

\bibitem[Unger \etal\ 1987]{ung}Unger, S. W., Lawrence, A., Wilson, A. S.,
Elvis, M., Wright, A. E. 1987, MNRAS, 288, 521

\bibitem[Terashima \etal\ 1994]{ser}Terashima, Y., Serlemitsos, P. J., Kunieda, H.,
and Iwasawa, K. 1994 , in `New Horizon of X-Ray Astronomy - First results from ASCA',
F. Makino and T. Ohashi eds., (Tokyo: Universal Academy Press), p. 523

\bibitem[Tully, Mould \& Aaronson 1982]{tma82}
Tully, B.R, Mould, J. R. \& Aaronson, M. 1982, \apj, 2557, 527.


\bibitem[Trinchieri \& Fabbiano 1985]{tf85}Trinchieri, G., Fabbiano, G. 1985, \apj, 296, 447

%\bibitem[Trinchieri \& Fabbiano 1991]{tf90}Trinchieri, G., Fabbiano, G. 1991, \apj, 382, 82


\bibitem[Trinchieri, Fabbiano \& Bandiera, 1989]{tfb89}
Trinchieri, G., Fabbiano, G., Bandiera, R. 1989, \apj, 342, 759
(TFB).

\bibitem[Young 1990]{y90}Young, J.S. in {\it Windows on Galaxies}, eds. G. Fabbiano, 
J.S. Gallagher, A. Renzini, p. 213, Dordrecht: Kluwer. 


\bibitem[Watson \etal\ 1984]{wsg84}Watson, M.G., Stanger, V., Griffiths, R.E. 1984, \apj, 286, 144



%\bibitem[Whitmore 1984]{whi84}Whitmore, B.C. 1984, \apj, 278, 61.

\bibitem[Wrobel \& Heeschen 1991]{wro}Wrobel, J.M., Heeschen, D.S. 1991, \aj, 101, 148







\end{thebibliography}
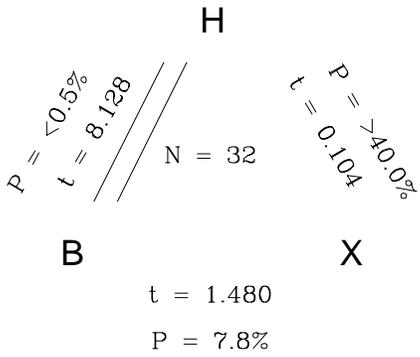
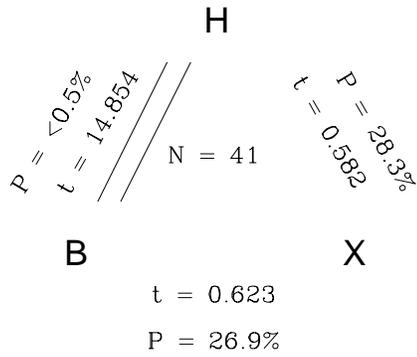
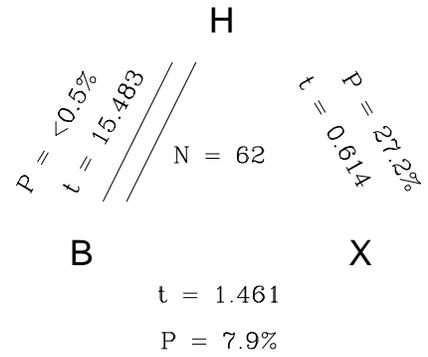
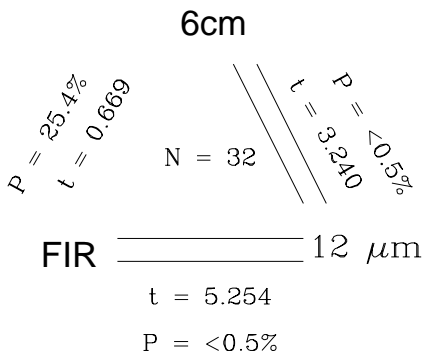
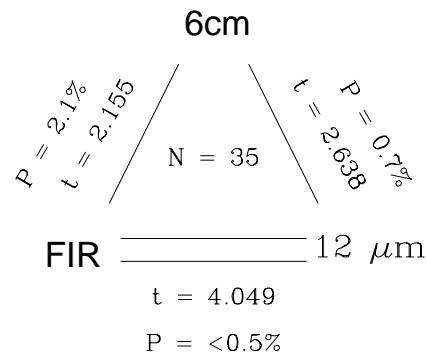
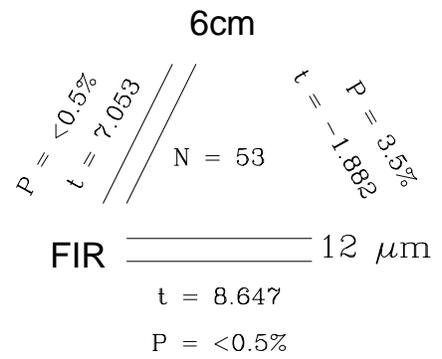
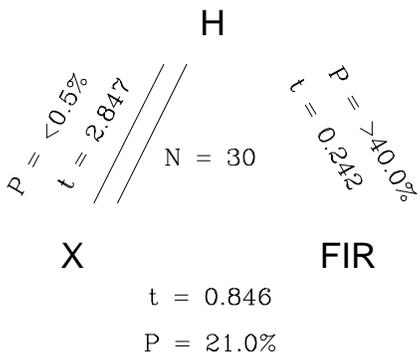
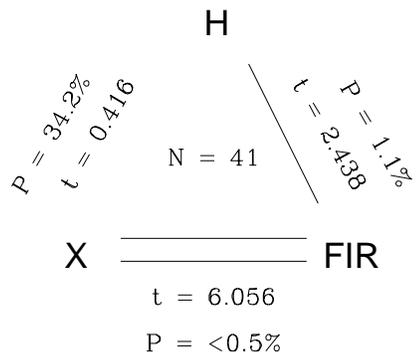
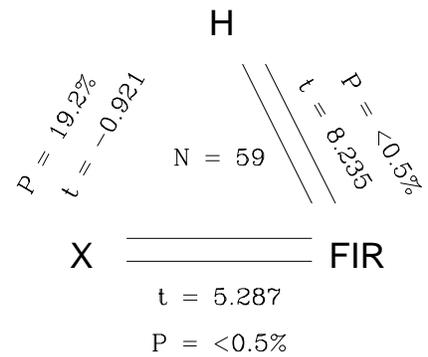
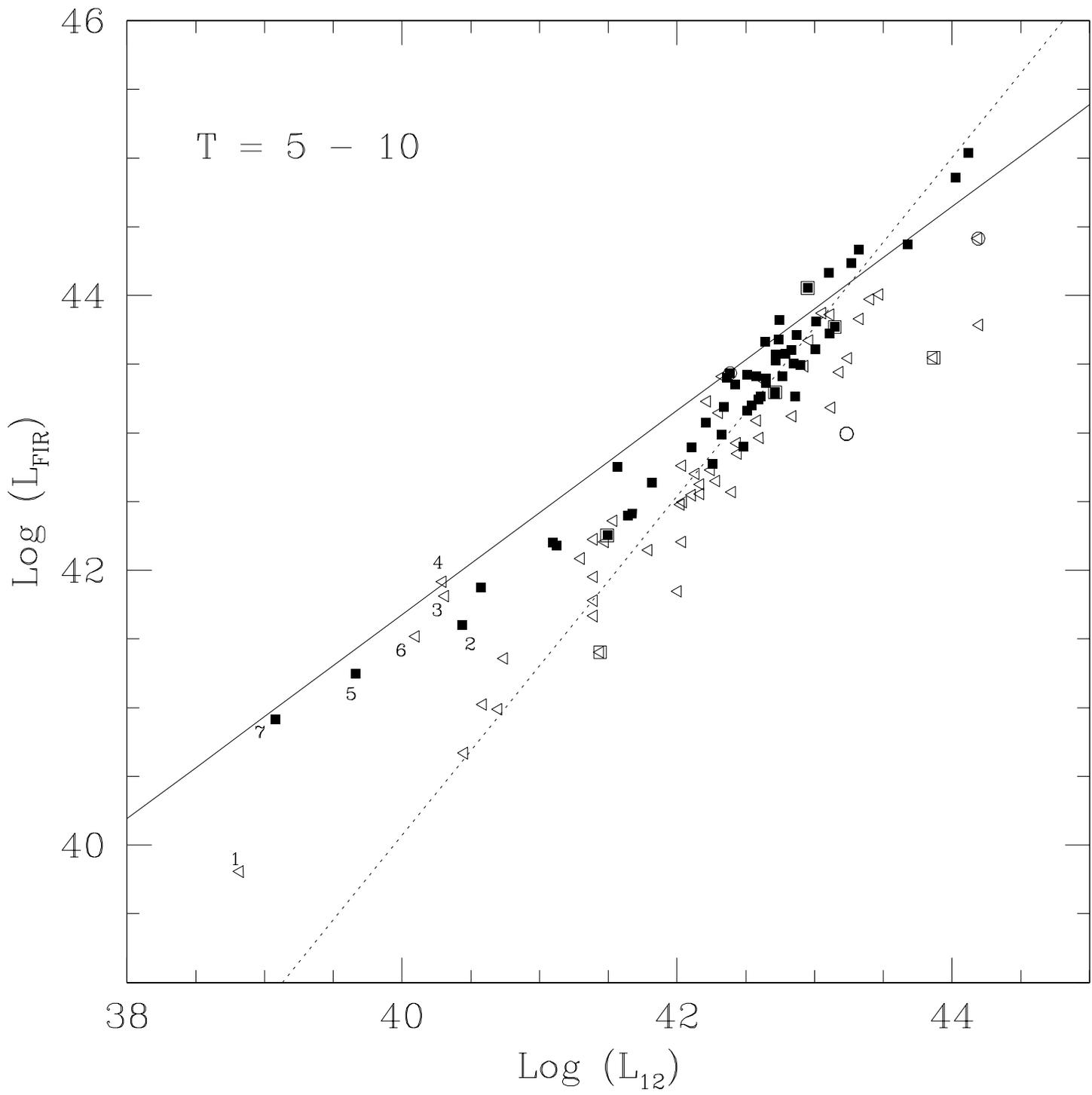
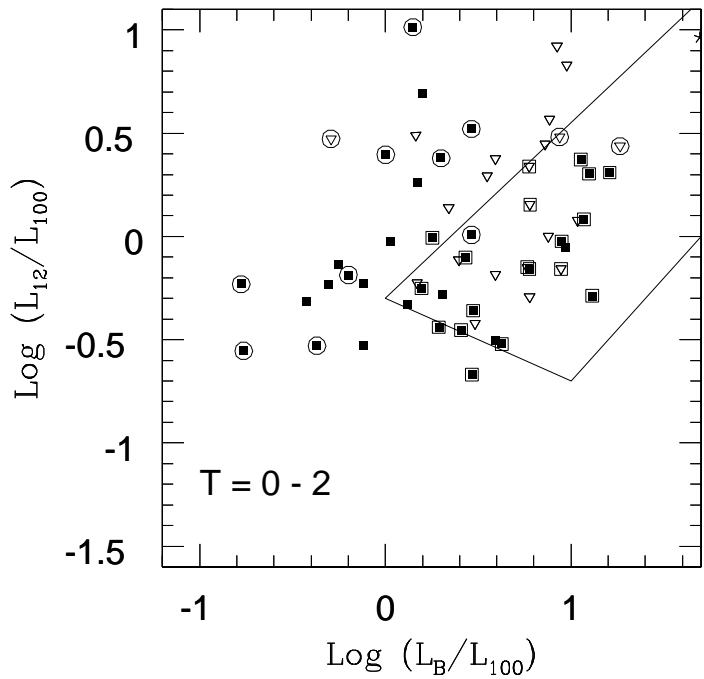
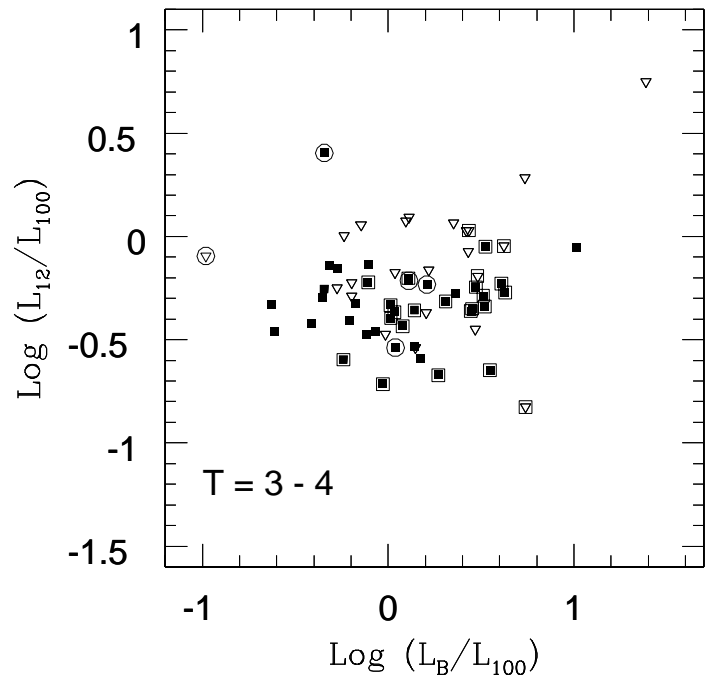
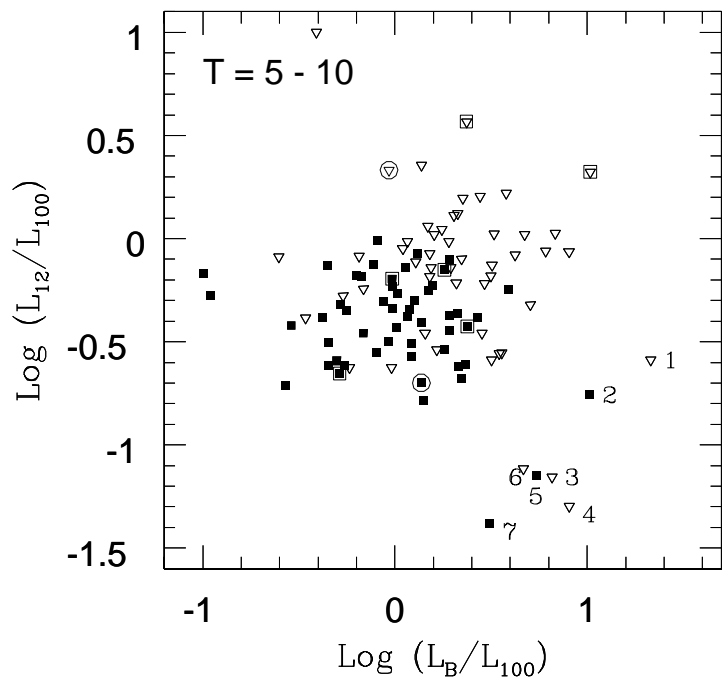
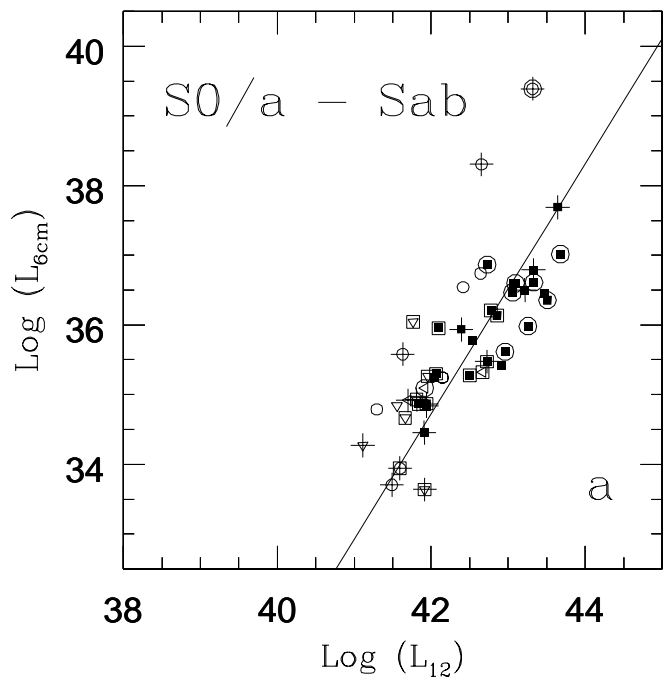
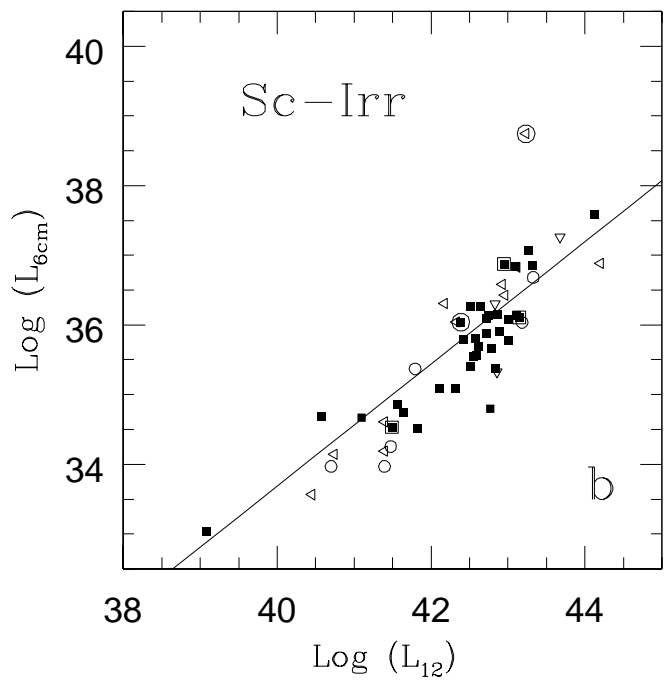
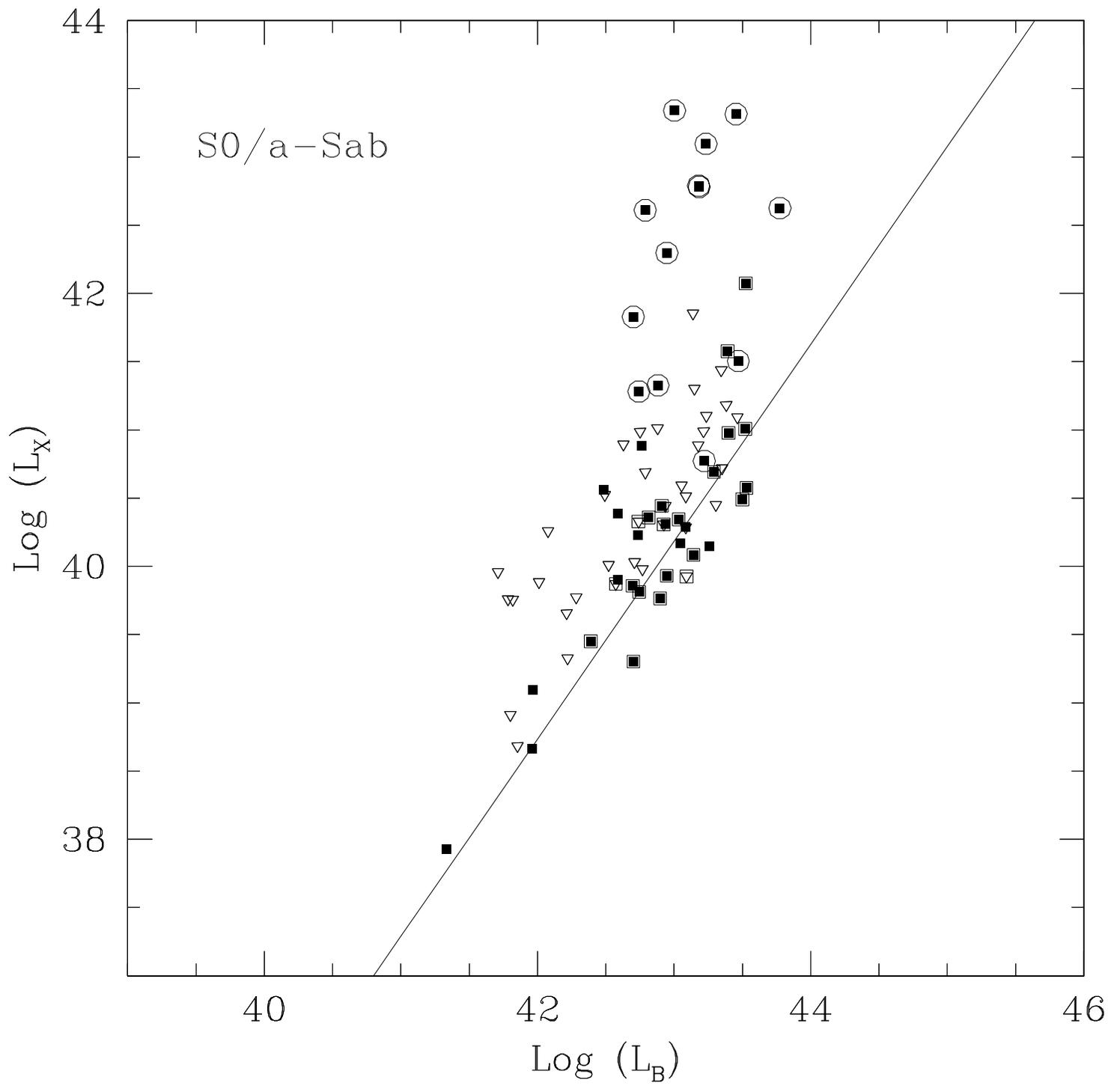
\end{document}